\documentclass[letterpaper,10pt,conference]{ieeeconf}
\pdfoutput=1

\IEEEoverridecommandlockouts  % This command is only needed if you want to use the \thanks command
\usepackage{graphicx}
\usepackage{newtxtext}
\usepackage{newtxmath}
\usepackage{bm}
\usepackage{amsmath}
\usepackage{booktabs}
\usepackage{tablefootnote}
\usepackage{xfrac}
\usepackage[ngerman,english]{babel}
\useshorthands{"}
\addto\extrasenglish{\languageshorthands{ngerman}}
\usepackage[hidelinks]{hyperref}
\usepackage{eso-pic}

\graphicspath{{./gfx/}{./gfx_gen/}}

\newcommand{\D}{\displaystyle}

\newcommand{\Vector}[1]{\bm{#1}}  % for vectors
\newcommand{\Matrix}[1]{\bm{#1}}  % for matrices
\newcommand{\Transpose}{\mathrm{T}}  % alternatives for transpose symbol: \mathrm{T}, \intercal, \top
\newcommand{\refEq}[1]{(\ref{#1})}               % equation
\newcommand{\refFig}[1]{Fig.~\ref{#1}}           % figure
\newcommand{\refSec}[1]{Section~\ref{#1}}        % section
\newcommand{\refSecBegin}[1]{Section~\ref{#1}}   % sec. at beginning of sentence
\newcommand{\refTable}[1]{table~\ref{#1}}        % table
\newcommand{\refTableBegin}[1]{Table~\ref{#1}}   % table at beginning of sentence

\title{\LARGE \bf
A Minimum-Footprint Implementation of Discrete-Time ADRC
}

\author{Gernot Herbst\ \href{https://orcid.org/0000-0002-4638-5378}%
    {\raisebox{-0.3pt}{\includegraphics[height=9pt]{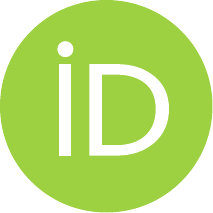}}}%
    \thanks{Gernot Herbst is with University of Applied Sciences Zwickau, Germany {\tt\small gernot.herbst@fh-zwickau.de}}%
}

\hypersetup{%
    pdfauthor={Gernot Herbst},
    pdftitle={A Minimum-Footprint Implementation of Discrete-Time ADRC},
    pdfcreator={},pdfproducer={}
}

\begin{document}

\AddToShipoutPicture*{%
  \AtPageUpperLeft{%
    \setlength\unitlength{1cm}%
    \put(0,-0.5){\begin{minipage}[t]{\paperwidth}
    \footnotesize\centering\textcolor{black!50}{%
    This is a preprint of an article accepted for publication at the 2021 European Control Conference (ECC21).\\%
    The final article is available at \textcolor{blue!60}{\href{https://doi.org/10.23919/ecc54610.2021.9655120}{https://doi.org/10.23919/ecc54610.2021.9655120}}. This revision of the preprint includes post-publication corrections.%
    }\end{minipage}}%
  }
}

\maketitle
\thispagestyle{empty}
\pagestyle{empty}

% -----------------------------------------------------------------------------
% -----------------------------------------------------------------------------

\begin{abstract}
To foster the adoption of active disturbance rejection control (ADRC) and support its deployment even on low-cost embedded systems, this article introduces the most efficient implementation of linear discrete-time ADRC to date. While maintaining all features and the exact performance characteristics of the state-space form, computational efforts and storage requirements are reduced to a minimum compared to all existing implementations. This opens up new possibilities to use ADRC in applications with high sample rates, tight timing constraints, or low computational power.
\end{abstract}

% -----------------------------------------------------------------------------
% Section: Introduction
% -----------------------------------------------------------------------------

\section{Introduction}

Active disturbance rejection control (ADRC), introduced by \cite{Han:2009} and streamlined by \cite{Gao:2003}, is a highly practical general purpose controller that has already found its way into numerous application domains \cite{Zheng:2010,Zheng:2018}. Conquering market share from PID type controllers, however, is not an easy feat and demands improvements in many areas. A lot has already been achieved: making ADRC easier to tune than PID \cite{Gao:2003,Herbst:2020}, providing a high-performance discrete-time form \cite{Miklosovic:2006}, equipping it with features for industrial practice \cite{Madonski:2015,Herbst:2016a}, enhancing the observer for improved disturbance rejection \cite{Madonski:2015b}, and lowering the barriers of understanding ADRC from a traditional control perspective by offering a comprehensive frequency-domain view \cite{Herbst:Preprint2020}.

Deployment on low-cost microprocessors, which are usually firmly in the grip of PID controllers, requires an efficient discrete-time software implementation of linear ADRC. This is the scope of this article. To the best of the author's knowledge, \cite{Herbst:Preprint2020} and \cite{Xu:2014} are---despite the multitude of recent research work in the domain of ADRC---the only publications addressing this topic so far. However, \cite{Xu:2014} uses Euler discretization, which was shown to be inferior to zero-order hold (ZOH) discretization in \cite{Miklosovic:2006}; while the approach of \cite{Herbst:Preprint2020} uses ZOH but does not support arbitrary limitation of the controller output signal, e.\,g.\ with a rate limiter, which can be a desirable feature in practice \cite{Herbst:2016a}.

This work presents the most efficient implementation of ADRC to date, at the same time keeping all features and exactly maintaining the performance delivered by ZOH discretization of linear ADRC based on the current observer approach, as recommended by \cite{Miklosovic:2006}.

\refSecBegin{sec:ADRC} of this paper summarizes the state-space form of discrete-time ADRC. \refSecBegin{sec:FBTF-ADRC} then introduces a novel transfer function representation that enables the efficient implementation presented in \refSec{sec:Implementation}. A comparison of the computational costs is given in \refSec{sec:Comparison}, followed by some practical issues in \refSec{sec:Initialization} and an example in \refSec{sec:Example}.

% -----------------------------------------------------------------------------
% Section: ADRC
% -----------------------------------------------------------------------------

\section{Discrete-Time Linear ADRC}
\label{sec:ADRC}

\begin{figure}[t]
    \centering%
    \includegraphics[width=\linewidth]{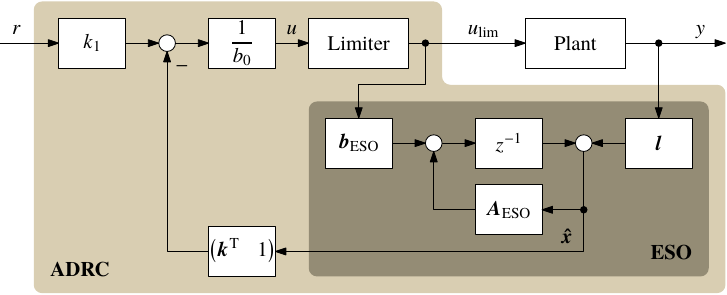}%
    \caption{Control loop with discrete-time ADRC. The controller output limitation can be a simple magnitude limitation, or include higher-order contraints. A magnitude- and rate-limiting variant can be found in \refFig{fig:Limiter_Rate}.}
    \label{fig:ADRC_ControlLoop}
\end{figure}

Aim of this section is to provide an introduction to dis"-crete"=time linear ADRC as briefly as possible. A  detailed introduction can be found in \cite{Herbst:2013}.

% -----------------------------------------------------------------------------

\subsection{Observer}
\label{sec:ADRC_Observer}

The core of linear ADRC is formed by a Luenberger observer for an $n^\mathrm{th}$-order integrator chain plant model with input gain $b_0$, extended by an unknown constant input disturbance, the so-called \emph{total disturbance}. Accordingly, the observer is also being called \emph{extended state observer} (ESO):
\begin{gather}
\Vector{\dot{\hat{x}}}(t)
=
\Matrix{A} \cdot \Vector{\hat{x}}(t)
+
\Vector{b} \cdot u(t)
+
\Vector{l} \cdot \left( y(t) - \Vector{c}^\Transpose \cdot  \Vector{\hat{x}}(t) \right)
\label{eqn:ADRC_Observer}
\\
\text{with}\quad
\Matrix{A} =
\begin{pmatrix}
\Matrix{0}^{n \times 1}  &  \Matrix{I}^{n \times n}
\\
0  &  \Matrix{0}^{1 \times n}
\end{pmatrix}
,\quad
\Vector{b} =
\begin{pmatrix}
\Matrix{0}^{(n-1) \times 1}
\\
b_0
\\
0
\end{pmatrix}
,
\notag
\\
\Matrix{c}^\Transpose =
\begin{pmatrix}
1  &  \Matrix{0}^{1 \times n}
\end{pmatrix}
,\quad
\Vector{l} =
\begin{pmatrix}
l_1  \\  \vdots  \\  l_{n+1}
\end{pmatrix}
,\quad
\Vector{\hat{x}} =
\begin{pmatrix}
\hat{x}_1  \\  \vdots  \\  \hat{x}_{n+1}
\end{pmatrix}
.
\notag
\end{gather}

For a discrete-time implementation of \refEq{eqn:ADRC_Observer}, \cite{Miklosovic:2006} recommends using zero-order hold (ZOH) discretization to transform $\Matrix{A}$, $\Vector{b}$, and $\Vector{c}^\Transpose$ into their discrete-time counterparts (with a sampling interval $T$):
\begin{gather}
\Matrix{A}_\mathrm{d} = \Matrix{I} + \sum_{i = 1}^\infty \D\frac{\Matrix{A}^i T^i}{i!}
,\ \
\Vector{b}_\mathrm{d} = \left( \sum_{i = 1}^\infty \D\frac{\Matrix{A}^{i-1} T^i}{i!} \right) \cdot \Vector{b}
,\ \
\Vector{c}^\Transpose_\mathrm{d} = \Vector{c}^\Transpose
.
\notag
%\label{eqn:ADRC_Discrete_Observer_ZOH}
\end{gather}

In \cite{Miklosovic:2006} it was shown as well that discrete-time ADRC should employ the current observer instead of the predictive observer approach, since the former bases the estimation of $\Vector{\hat{x}}(k)$ on the most recent measurement $y(k)$. In this article, the abbrevations $\Matrix{A}_\mathrm{ESO}$ and $\Vector{b}_\mathrm{ESO}$ introduced in \cite{Herbst:2013} will be used in the observer equations \refEq{eqn:ADRC_Discrete_ESO}.

\begin{gather}
\Vector{\hat{x}}(k) = \Matrix{A}_\mathrm{ESO} \cdot \Vector{\hat{x}}(k-1) + \Vector{b}_\mathrm{ESO} \cdot u(k-1) + \Vector{l} \cdot y(k)
\label{eqn:ADRC_Discrete_ESO}
\\
\text{with}\quad
\Matrix{A}_\mathrm{ESO} = \Matrix{A}_\mathrm{d} - \Vector{l} \cdot \Vector{c}^\Transpose_\mathrm{d} \cdot \Matrix{A}_\mathrm{d}
,\quad
\Vector{b}_\mathrm{ESO} = \Vector{b}_\mathrm{d} - \Vector{l} \cdot \Vector{c}^\Transpose_\mathrm{d} \cdot \Vector{b}_\mathrm{d}
.
\notag
\end{gather}

% -----------------------------------------------------------------------------

\subsection{Control Law}
\label{sec:ADRC_Controller}

Linear ADRC employs an ordinary $n^\mathrm{th}$-order state feedback controller; only enhanced by feeding back the estimated total disturbance $\hat{x}_{n+1}$, and a common factor $\frac{1}{b_0}$ to neutralize the plant gain $b_0$. Since the observer has been discretized already, the discrete-time control law can be trivially obtained from the continuous-time case by replacing time $t$ with the sampling instant $k$:
\begin{gather}
u(k) = \frac{1}{b_0} \cdot \left( k_1 \cdot r(k) -
\begin{pmatrix}
\Vector{k}^\Transpose  &  1
\end{pmatrix}
\cdot \Vector{\hat{x}}(k)
\right)
\label{eqn:ADRC_Discrete_Controller}
\\
\text{with}\quad
\Vector{k}^\Transpose =
\begin{pmatrix}
k_1  &  \cdots  &  k_n
\end{pmatrix}
.
\notag
\end{gather}

% -----------------------------------------------------------------------------

\subsection{Controller and Observer Tuning}
\label{sec:ADRC_Tuning}

The benefit of feeding back the total disturbance and neutralizing the plant gain $b_0$ is that the state feedback controller gains $\Vector{k}^\Transpose$ can be tuned for a virtual plant behaving as a pure $n^\mathrm{th}$-order integrator chain, depending only on the desired closed-loop dynamics. The most popular approach is called \emph{bandwidth parameterization} \cite{Gao:2003}, placing all closed-loop poles at $-\omega_\mathrm{CL}$. This leads to the design equations:
\begin{equation}
k_i = \frac{n!}{(n-i+1)! \cdot (i-1)!} \cdot \omega_\mathrm{CL}^{n-i+1}
\quad \forall i = 1, ..., n
.
\label{eqn:CtrlDesign_K_PolePlacement_Param}
\end{equation}

Following the bandwidth tuning approach, the observer of an $n^\mathrm{th}$-order discrete-time ADRC is  tuned by placing all eigenvalues of $\Matrix{A}_\mathrm{ESO}$ at a common location $z_\mathrm{ESO}$:
\begin{gather}
\det\left( z \Matrix{I} - \Matrix{A}_\mathrm{ESO} \right)
\stackrel{!}{=} \left( z - z_\mathrm{ESO} \right)^{n+1}
\label{eqn:ADRC_Design_L_Approach_Discrete}
\\
\text{with}\quad
z_\mathrm{ESO} = \mathrm{e}^{- k_\mathrm{ESO} \cdot \omega_\mathrm{CL} \cdot T}
,
\notag
\end{gather}
where $k_\mathrm{ESO}$ is a relative factor defining the observer bandwidth relative to the desired closed-loop bandwidth $\omega_\mathrm{CL}$.

In summary, discrete-time ADRC consists of a controller and observer given in equations \refEq{eqn:ADRC_Discrete_Controller} and \refEq{eqn:ADRC_Discrete_ESO}, which are illustrated in \refFig{fig:ADRC_ControlLoop}; and is being tuned by solving equations \refEq{eqn:CtrlDesign_K_PolePlacement_Param} and \refEq{eqn:ADRC_Design_L_Approach_Discrete}.

\begin{figure}[t]
    \centering%
    \includegraphics[]{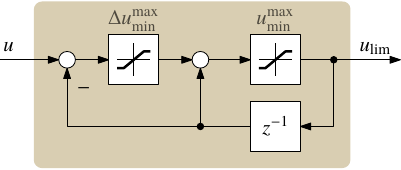}%
    \caption{Discrete-time magnitude and rate limitation block from \cite{Herbst:2016a} that can be used in an ADRC control loop as given in \refFig{fig:ADRC_ControlLoop}.}
    \label{fig:Limiter_Rate}
\end{figure}

% -----------------------------------------------------------------------------
% Section: FBTF-ADRC
% -----------------------------------------------------------------------------

\section{Feedback Transfer Function-Based ADRC}
\label{sec:FBTF-ADRC}

\begin{figure}[t]
    \centering%
    \includegraphics[width=\linewidth]{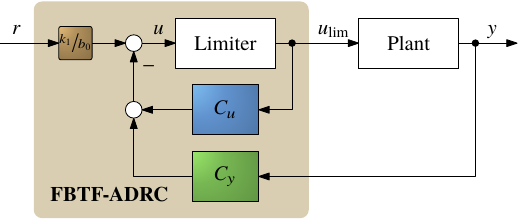}%
    \caption{Control loop with a discrete-time implementation of ADRC based on feedback transfer functions (FBTF-ADRC).}
    \label{fig:ADRC_ControlLoop_FBTF}
\end{figure}

From \refFig{fig:ADRC_ControlLoop} it becomes apparent that the controller dynamics of ADRC are located in the feedback paths from the measured plant output $y$ and the previous limited controller output $u_\mathrm{lim}$ back to the unlimited controller output $u$. In this article, we therefore propose to implement discrete-time ADRC using transfer functions in the feedback path, as illustrated in \refFig{fig:ADRC_ControlLoop_FBTF} (FBTF-ADRC). This marks a departure from other controller implementations based on transfer functions. Usually, a comparable two-degrees-of-freedom (2DOF) controller is being implemented using a reference signal filter and a controller transfer function acting on the control error.

To obtain these two feedback transfer functions $C_u(z)$ and $C_y(z)$ from \refFig{fig:ADRC_ControlLoop_FBTF}, put \refEq{eqn:ADRC_Discrete_ESO} into \refEq{eqn:ADRC_Discrete_Controller} in $z$-domain:
\begin{gather*}
u(z) =
\frac{k_1}{b_0} \cdot r(z)
- C_u(z) \cdot u_\mathrm{lim}(z)
- C_y(z) \cdot y(z)
\\
\text{with}\quad
C_u(z) =
\frac{z^{-1}}{b_0}
\cdot
\begin{pmatrix}
\Vector{k}^\Transpose  &  1
\end{pmatrix}
\cdot
\left( \Matrix{I} - z^{-1} \Matrix{A}_\mathrm{ESO} \right)^{-1}
\cdot
\Vector{b}_\mathrm{ESO}
\\
\text{and}\quad
C_y(z)
=
\frac{1}{b_0}
\cdot
\begin{pmatrix}
\Vector{k}^\Transpose  &  1
\end{pmatrix}
\cdot
\left( \Matrix{I} - z^{-1} \Matrix{A}_\mathrm{ESO} \right)^{-1}
\cdot
\Vector{l}
.
\end{gather*}

When applying bandwidth tuning as described in \refSec{sec:ADRC_Tuning} to the extended state observer (placing all poles of $\Matrix{A}_\mathrm{ESO}$ at $z_\mathrm{ESO}$, cf.\ \refEq{eqn:ADRC_Design_L_Approach_Discrete}), one obtains the following denominators for the two transfer functions:
\begin{gather*}
C_u(z) =
\frac{z^{-1}}{b_0}
\cdot
\frac{
\begin{pmatrix}
\Vector{k}^\Transpose  &  1
\end{pmatrix}
\cdot
\operatorname{adj} \left( \Matrix{I} - z^{-1} \Matrix{A}_\mathrm{ESO} \right)
\cdot
\Vector{b}_\mathrm{ESO}
}{ \left( 1 - z^{-1} z_\mathrm{ESO} \right)^{n+1} }
\\
C_y(z)
=
\frac{1}{b_0}
\cdot
\frac{
\begin{pmatrix}
\Vector{k}^\Transpose  &  1
\end{pmatrix}
\cdot
\operatorname{adj} \left( \Matrix{I} - z^{-1} \Matrix{A}_\mathrm{ESO} \right)
\cdot
\Vector{l}
}{ \left( 1 - z^{-1} z_\mathrm{ESO} \right)^{n+1} }
\end{gather*}

Using bandwidth tuning for the controller as well, cf.\ \refEq{eqn:CtrlDesign_K_PolePlacement_Param}, we can write the two feedback transfer functions as follows, forming the basis for a minimum-footprint implementation:
\begin{gather}
\label{eqn:CuCy}
C_u(z)
=
z^{-1} \cdot
\frac{ \D\sum_{i = 0}^{n} \beta_i z^{-i } }{ 1 + \D\sum_{i = 1}^{n+1} \alpha_i z^{-i} }
\\
C_y(z)
=
\frac{ \D\sum_{i = 0}^{n} \gamma_i z^{-i} }{ 1 + \D\sum_{i = 1}^{n+1} \alpha_i z^{-i} }
\end{gather}

For first- and second-order ADRC, the resulting coefficients required for the proposed controller implementation ($\alpha_{1,\ldots,n+1}$, $\beta_{0,\ldots,n}$, $\gamma_{0,\ldots,n}$, $\frac{k_1}{b_0}$) are provided in \refTable{table:TF_Parameters}.

\begin{table*}
\begin{center}
\caption{Coefficients of the proposed discrete-time feedback transfer function implementation of first- and second-order ADRC, given in terms of the bandwidth tuning parameters $\omega_\mathrm{CL}$ (desired closed-loop bandwidth) and $z_\mathrm{ESO} = \mathrm{e}^{- k_\mathrm{ESO} \cdot \omega_\mathrm{CL} \cdot T}$ (observer pole locations in $z$-domain, using the observer bandwidth factor $k_\mathrm{ESO}$); as well as $b_0$ (gain parameter of the plant model) and $T$ (sample time of the discrete-time implementation).}
\label{table:TF_Parameters}
\begin{tabular*}{\linewidth}{@{\extracolsep\fill}rll@{\extracolsep\fill}}
    \toprule
    \textbf{Coeff.}  &  \textbf{Value for first-order ADRC}  &  \textbf{Value for second-order ADRC}    \\
    \midrule
    $\alpha_1$
    &   $-2 z_\mathrm{ESO}$
    &   $-3 z_\mathrm{ESO}$
    \\[1.2em]
    $\alpha_2$
    &   $z_\mathrm{ESO}^2$
    &   $3 z_\mathrm{ESO}^2$
    \\[1.2em]
    $\alpha_3$
    &   ---
    &   $-z_\mathrm{ESO}^3$
    \\[1.2em]
    $\beta_0$
    &   $\D T \omega_\mathrm{CL} z_\mathrm{ESO}^2 - \left( 1 - z_\mathrm{ESO} \right)^2$
    &   $\D \frac{1}{2} \cdot \left[ -T \omega_\mathrm{CL} z_\mathrm{ESO}^3 \cdot \left( 4 - T \omega_\mathrm{CL} \right) + T \omega_\mathrm{CL} \cdot \left( 1 + z_\mathrm{ESO} \right)^3 - \left( 1 - z_\mathrm{ESO} \right)^3 \right]$
    \\[1.2em]
    $\beta_1$
    &   $\D -T \omega_\mathrm{CL} z_\mathrm{ESO}^2$
    &   $\D \frac{1}{2} \cdot \left[ -T \omega_\mathrm{CL} \cdot \left( 1 + z_\mathrm{ESO} \right)^3 - \left( 1 - z_\mathrm{ESO} \right)^3 \right]$
    \\[1.2em]
    $\beta_2$
    &  ---
    &   $\D \frac{1}{2} \cdot \left[ T \omega_\mathrm{CL} z_\mathrm{ESO}^3 \cdot \left( 4 - T \omega_\mathrm{CL} \right) \right]$
    \\[1.2em]
    $\gamma_0$
    &   $\D \frac{1}{b_0 T} \cdot \left[ T \omega_\mathrm{CL} \cdot \left( 1 - z_\mathrm{ESO}^2 \right) +  \left( 1 - z_\mathrm{ESO} \right)^2 \right]$
    &   $\D \frac{1}{b_0 T^2} \cdot \left[ T^2 \omega_\mathrm{CL}^2 \cdot \left( 1 - z_\mathrm{ESO}^3 \right) + 3 T \omega_\mathrm{CL} \cdot \left( 1 - z_\mathrm{ESO} - z_\mathrm{ESO}^2 + z_\mathrm{ESO}^3 \right) + \left( 1 - z_\mathrm{ESO} \right)^3 \right]$
    \\[1.2em]
    $\gamma_1$
    &   $\D \frac{1}{b_0 T} \cdot \left[ 2 T \omega_\mathrm{CL} \cdot \left( z_\mathrm{ESO}^2 - z_\mathrm{ESO} \right) -  \left( 1 - z_\mathrm{ESO} \right)^2 \right]$
    &   $\D \frac{1}{b_0 T^2} \cdot \left[ 3 T^2 \omega_\mathrm{CL}^2 \cdot \left( -z_\mathrm{ESO} + z_\mathrm{ESO}^3 \right) + 4 T \omega_\mathrm{CL} \cdot \left( -1 + 3 z_\mathrm{ESO}^2 - 2 z_\mathrm{ESO}^3 \right) - 2 \cdot \left( 1 - z_\mathrm{ESO} \right)^3 \right]$
    \\[1.2em]
    $\gamma_2$
    &  ---
    &   $\D \frac{1}{b_0 T^2} \cdot \left[ 3 T^2 \omega_\mathrm{CL}^2 \cdot \left( z_\mathrm{ESO}^2 - z_\mathrm{ESO}^3 \right) + T \omega_\mathrm{CL} \cdot \left( 1 +  3 z_\mathrm{ESO} - 9 z_\mathrm{ESO}^2 + 5 z_\mathrm{ESO}^3 \right) + \left( 1 - z_\mathrm{ESO} \right)^3 \right]$
    \\[1.2em]
    $\D\frac{k_1}{b_0}$
    &   $\D \frac{\omega_\mathrm{CL}}{b_0}$
    &   $\D \frac{\omega_\mathrm{CL}^2}{b_0}$
    \\[0.9em]
    \bottomrule
\end{tabular*}
\end{center}
\end{table*}

% -----------------------------------------------------------------------------
% Section: Implementation
% -----------------------------------------------------------------------------

\section{Efficient Implementation}
\label{sec:Implementation}

In this section, a discrete-time ADRC implementation with focus on minimum computational footprint (in terms of required storage variables and mathematical operations) shall be presented. As a starting point, we write the unlimited controller output $u$ from \refFig{fig:ADRC_ControlLoop_FBTF} in time domain as follows,
\begin{equation}
u(k) = \frac{k_1}{b_0} \cdot r(k) - c(k)
,
\end{equation}
where $c$ is the combined output of the feedback transfer functions $C_u$ and $C_y$ in time domain, which, due to their identical denominator, can be written and implemented as a superposition:
\begin{equation}
\begin{aligned}
c(k) = \sum_{i = 0}^{n} \Big[ & -\alpha_{i+1} \cdot c(k-i-1)
\\
                               & + \beta_i \cdot u_\mathrm{lim}(k-i-1) + \gamma_i \cdot y(k-i) \Big].
\end{aligned}
\label{eqn:Imp_c_of_k}
\end{equation}

\begin{figure}[tb]
    \centering%
    \includegraphics[]{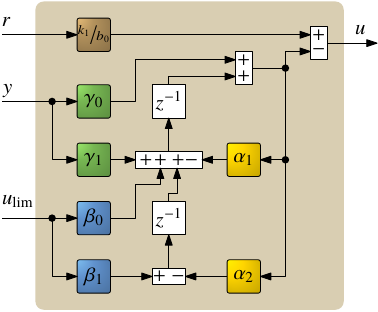}%
    \caption{Proposed minimum-footprint implementation of first-order ADRC based on feedback transfer functions, cf.\ \refFig{fig:ADRC_ControlLoop_FBTF}, excluding the output limitation. The limited controller output value has to be provided on the input $u_\mathrm{lim}$. Values for all coefficients (colored boxes) are provided in \refTable{table:TF_Parameters}.}
    \label{fig:ADRC_FBTF_1}
\end{figure}

\begin{figure}[tb]
    \centering%
    \includegraphics[]{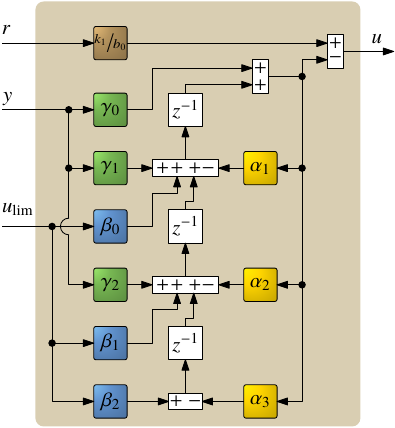}%
    \caption{Minimum-footprint implementation of second-order ADRC. Values for all coefficients (colored boxes) are provided in \refTable{table:TF_Parameters}.}
    \label{fig:ADRC_FBTF_2}
\end{figure}

Using a common (factored-out) denominator will save multiplications in the control law implementation on the target hardware. The other ingredient for an efficient implementation proposed here is the use of a transposed direct form II \cite{Oppenheim:2010} for \refEq{eqn:Imp_c_of_k}. This reduces the number of necessary storage variables to a minimum. For the first- and second-order cases of discrete-time linear ADRC, the resulting implementation is depicted in \refFig{fig:ADRC_FBTF_1} and \refFig{fig:ADRC_FBTF_2}, respectively.

Let $x_1, \ldots, x_{n+1}$ be the value of the storage variables (unit delays) in \refFig{fig:ADRC_FBTF_1} and \refFig{fig:ADRC_FBTF_2}, numbered in increasing order from top to bottom. We can now give the detailed equations for updating outputs and internal states of the ADRC implementation proposed in this article:
\begin{enumerate}
\item
Compute the combined transfer function output $c(k)$:\\
$\D c(k) = \gamma_0 \cdot y(k) + x_1(k-1)$.

\item
Compute the unlimited controller output $u(k)$:\\
$\D u(k) = \frac{k_1}{b_0} \cdot r(k) - c(k)$.

\item
Limit the controller output $u(k)$ to obtain $u_\mathrm{lim}(k)$, e.\,g.\ using a rate limitation block as given in \refFig{fig:Limiter_Rate}.

\item
Update the upper $n$ storage variables $x_i$ (unit delays) from top to bottom, i.\,e.\ $\forall i = 1, \ldots, n$:\\
$\D x_i(k) = x_{i+1}(k-1) - \alpha_i \cdot c(k) + \beta_{i-1} \cdot u_\mathrm{lim}(k) + \gamma_i \cdot y(k)$.

\item
Update the storage variable $x_{n+1}$:\\
$\D x_{n+1}(k) = -\alpha_{n+1} \cdot c(k) + \beta_{n} \cdot u_\mathrm{lim}(k)$.
\end{enumerate}

% -----------------------------------------------------------------------------
% Section: Comparison
% -----------------------------------------------------------------------------

\section{Comparison}
\label{sec:Comparison}

To assess the computational footprint of the proposed ADRC implementation, a comparison regarding the number of required storage variables, multplications, and additions is being made here. \refTableBegin{table:Costs} lists the computational costs for discrete-time ADRC of orders one, two, and $n$ (i.\,e.\ the general case). In \refTable{table:Costs}, only exact implementations of discrete-time ADRC based on ZOH discretization and the current observer form (the best performing approach recommended by \cite{Miklosovic:2006}) are being compared:

\begin{enumerate}
\item
The discrete-time state-space implementation given in \refEq{eqn:ADRC_Discrete_ESO} and \refEq{eqn:ADRC_Discrete_Controller}, as described by \cite{Herbst:2013}. The state-space approach features a minimum number of required storage variables ($n+1$ for an $n^\mathrm{th}$-order ADRC), but unfortunately, the number of required arithmetic operations scales quadratically with $n$.

\item
The implementation based on discrete-time transfer functions, as recently introduced in \cite{Herbst:Preprint2020}. At the cost of more required storage variables, this approach greatly reduces the number of required arithmetic operations, which only scales linearly with $n$. This, of course, becomes especially apparent for higher orders $n$. On the downside, this approach does not retain all features of the observer, therefore this implementation is not as capable of dealing with arbitrary control signal limitations as the original state-space form. The reason for this is that the limited control signal is not being fed back, since the observer implementation is moved and hidden in an error-based feedback transfer function.

\item
The implementation proposed in this article combines the best of both worlds (state space and transfer functions): The number of storage variables is equally low as in the state-space approach, and the number of arithmetic operations is as low as in \cite{Herbst:Preprint2020} for first-order ADRC, and even lower for $n \ge 2$. Since the feedback path of the limited controller output ($u_\mathrm{lim}$ in \refFig{fig:ADRC_ControlLoop_FBTF}) is retained, the ability of state-space ADRC to cope with control signal constraints such as rate limitation is being preserved.
\end{enumerate}

\begin{table}
\begin{center}
\caption{Comparison of computational costs (multiplications, additions, and storage variables; excluding costs of optional limiter) of different discrete-time implementations of ADRC of order 1, 2, and $n$. Best values are highlighted.}
\label{table:Costs}
\begin{tabular*}{\linewidth}{@{\extracolsep\fill}rlll@{\extracolsep\fill}}
%\begin{tabular}{rlll}
    \toprule
    \textbf{Order}  &  \textbf{State space \cite{Herbst:2013}}  &  \textbf{Transfer funct. \cite{Herbst:Preprint2020}\textsuperscript{$\dagger$}}  &  \textbf{Proposed}  \\
    \midrule
    1\textsuperscript{st}  &  11 mul.  &  \textbf{7} mul.  &  \textbf{7} mul.  \\
                           &  10 add.  &  \textbf{6} add.  &  \textbf{6} add.  \\
                           &  \textbf{2} var.  &  4 var.  &  \textbf{2} var.  \\
    \addlinespace[\aboverulesep] \cmidrule(r{1.5em}){2-4} \addlinespace[\belowrulesep]
    2\textsuperscript{nd}  &  19 mul.  &  11 mul.  &  \textbf{10} mul.  \\
                           &  18 add.  &  10 add. &  \textbf{9} add.  \\
                           &  \textbf{3} var.  &  6 var.  &  \textbf{3} var.  \\
    \addlinespace[\aboverulesep] \cmidrule(r{1.5em}){2-4} \addlinespace[\belowrulesep]
    $n$                    &  $(n^2 + 5n + 5)$ mul.  &  $(4n + 3)$ mul.  &  $(\bm{3n + 4})$ mul.  \\
                           &  $(n^2 + 5n + 4)$ add.  &  $(4n + 2)$ add.  &  $(\bm{3n + 3})$ add.  \\
                           &  $(\bm{n + 1})$ var.  &  $(2n + 2)$ var.  &  $(\bm{n + 1})$ var.  \\
    \bottomrule
    \addlinespace[\belowrulesep]
    \multicolumn{4}{l}{\textsuperscript{$\dagger$}\parbox[t]{0.93\linewidth}{\footnotesize{%
        Note that, for a fair comparison, the numbers of additions and variables given for the discrete-time transfer function approach from \cite{Herbst:Preprint2020} refer to a direct form II (transposed) implementation, as well.%
    }}}
\end{tabular*}
\end{center}
\end{table}

Note that the approach of \cite{Xu:2014}, the only other known ADRC implementation aiming at efficiency besides \cite{Herbst:Preprint2020}, was not included in \refTable{table:Costs} since it is based on an Euler discretization, which was shown to be inferior to ZOH in \cite{Miklosovic:2006}. Regarding arithmetic operations, it is on the same level as the proposed approach with $(3n + 4)$ multiplications and $(3n + 3)$ additions, but it requires $(3n + 5)$ variables compared to $(n + 1)$ in this article.

% -----------------------------------------------------------------------------
% Section: Initialization
% -----------------------------------------------------------------------------

\section{Controller Initialization}
\label{sec:Initialization}

An important issue in real-world control systems is to initialize the internal states before enabling the controller in order to realize bumpless transfers from manual mode or other controllers \cite{Herbst:2016a}. For the ADRC implementation proposed here, different strategies for initializing the internal states (storage variables $x_1, \ldots, x_{n+1}$, cf.\ \refSec{sec:Implementation}) are possible.

%NOTE: figure moved here from section "Example" for earlier placement in paper
\begin{figure*}
    \includegraphics{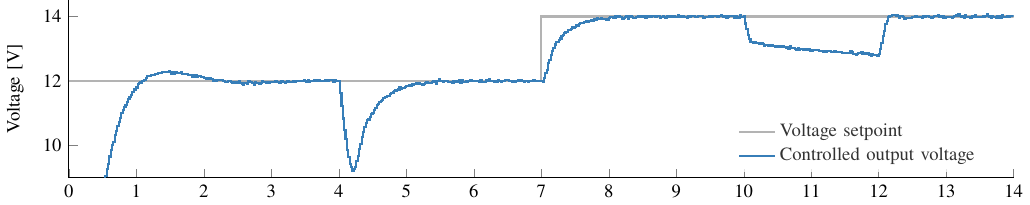}
    \includegraphics{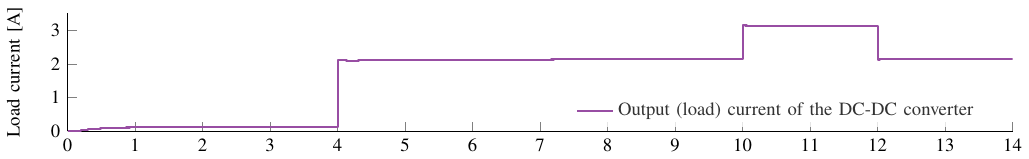}
    \includegraphics{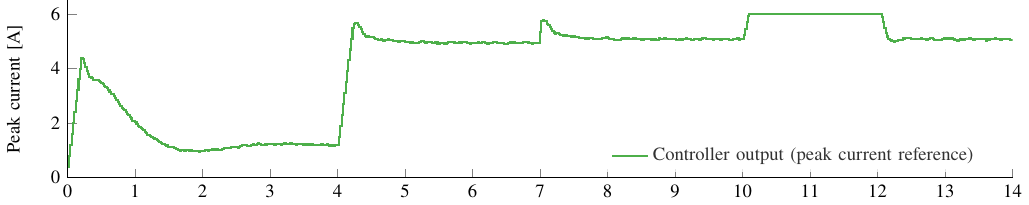}
    \includegraphics{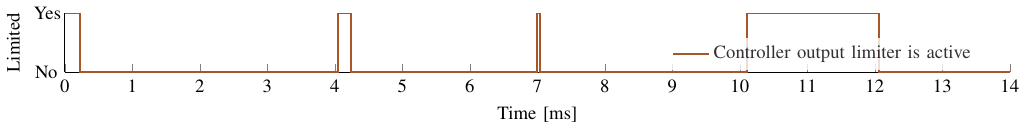}
    \caption{Simulation example: first-order discrete-time ADRC as an outer voltage controller for a peak current mode (PCM)-controlled step-down DC-DC converter. The controller output (green) is limited both in magnitude (max.\ $6\,\mathrm{A}$) and rate (max. $\pm 20\,\mathrm{A/ms}$), with the brown graph indicating if the output limitation is currently in action. Besides the voltage setpoint step at $t = 7\,\mathrm{ms}$ (grey), different load current steps (purple) are performed at $t = 4\,\mathrm{ms}$, $10\,\mathrm{ms}$, and $12\,\mathrm{ms}$. Within $t \in [10\,\mathrm{ms}, 12\,\mathrm{ms}]$, the load current is deliberately set high enough to drive the voltage controller into saturation. From the controller output (green) one can nicely observe the magnitude and rate limitation being in action, and the controller immediately recovering from being in saturation.}
    \label{fig:Example}
\end{figure*}

\paragraph{Tracking}
If there is enough time to let the internal states settle before enabling the controller, the initialization can simply be performed by running the control algorithm as described in \refSec{sec:Implementation} also while being disabled, with the following modification: Skip steps (2) and (3) of the control algorithm, instead let the output limitation track the externally given controller output $u^*$ such that $u_\mathrm{lim}(k) = u^*$.

\paragraph{Direct initialization}
For directly enabling the controller with correctly initialized states, the following steps have to be performed:
\begin{enumerate}
\item
Compute the combined transfer function output $c(k)$ from the current measurement $y(k)$ and the externally given initial controller output $u^*$:\\
$\D c(k) = \frac{k_1}{b_0} \cdot y(k) - u^*$.

\item
Initialize the output limitation block such that $u_\mathrm{lim}(k) = u^*$ holds.

\item
Initialize the storage variable $x_{n+1}$:\\
$\D x_{n+1}(k) = -\alpha_{n+1} \cdot c(k) + \beta_{n} \cdot u_\mathrm{lim}(k)$.

\item
Update the upper $n$ storage variables $x_i$ in reverse order, i.\,e.\ $\forall i = n, n-1, \ldots, 1$:\\
$\D x_i(k) = x_{i+1}(k) - \alpha_i \cdot c(k) + \beta_{i-1} \cdot u_\mathrm{lim}(k) + \gamma_i \cdot y(k)$.
\end{enumerate}

% -----------------------------------------------------------------------------
% Section: Example
% -----------------------------------------------------------------------------

\addtolength{\textheight}{-5.5cm}   % This command serves to balance the column lengths.

\section{Example}
\label{sec:Example}

To showcase some features of the proposed discrete-time ADRC implementation, a realistic simulation example shall be presented here. In this example, a first-order ADRC is being used as the outer voltage controller of a peak current mode (PCM)-controlled step-down DC-DC converter. For this topology, the outer voltage controller usually is a PI or ``Type 2'' controller (PI with additional low-pass filter) \cite{Suntio:2018}.

The converter parameters are: inductor $L = 33\,\mathrm{\upmu H}$, capacitor $C = 100\,\mathrm{\upmu F}$, switching and control frequency: $50\,\mathrm{kHz}$. To prevent subharmonic oscillations, the slope compensation of the inner peak current mode controller is set to $2\,\mathrm{A}$ per cycle \cite{Hallworth:2012}. The load of the converter is made of a purely ohmic base load with $R = 100\,\mathrm{\Omega}$, and a programmable constant current sink.

Controller design of a first-order ADRC for the PCM-controlled buck converter is extremely easy, the critical gain parameter $b_0$ can simply be set to the reciprocal of the output capacitance, i.\,e.\ $b_0 = 1/C = 10000$, cf.\  \cite{Herbst:2016a}. For demonstration purposes, the controller output limitation is set to a magnitude limit of $[0\,\mathrm{A}, 6\,\mathrm{A}]$ and a rate limit of $\pm 20\,\mathrm{A/ms}$. The closed-loop bandwidth is set to $\omega_\mathrm{CL} = 4000\,\mathrm{rad/s}$, corresponding to a settling time of $1\,\mathrm{ms}$ \cite{Herbst:2013}, and the relative observer bandwidth to $k_\mathrm{ESO} = 5$.

For a more realistic simulation, measurement noise is added to the controlled output voltage with $\sigma = 20\,\mathrm{mV}$. Additionally, a delay of one full sample cycle is inserted between measurement and updated controller output, which can be seen as a worst-case assumption \cite{Hallworth:2012}. The controller (as proposed in \refFig{fig:ADRC_FBTF_1}, plus rate limiter from \refFig{fig:Limiter_Rate}) is being simulated using single precision floating point, corresponding to a typical mainstream microcontroller implementation. The converter part is being simulated by a detailed circuit simulation, with the only simplification being ideal transistor and diode elements.

In the simulation scenario, the control loop is faced with startup from zero voltage, rapid load changes, a setpoint change, and being temporarily driven into saturation. The results are presented in \refFig{fig:Example}.

The application example presented here demonstrates both general favorable properties of linear ADRC as well as properties of the discrete-time implementation proposed in this article: The controller was very easy to tune, and the desired settling time is exactly visible in setpoint change in \refFig{fig:Example}. The presence of measurement noise does not deteriorate the control signal. Magnitude and rate constraints can easily be applied to the controller output signal, and thanks to the ``built-in'' anti-windup capability of ADRC, cf.\ \cite{Herbst:2016a}, no additional implementation or tuning effort is necessary regarding anti-windup measures. In \refFig{fig:Example} one can witness both magnitude and rate limitation in action, and watch the controller immediately recover from being in saturation. This is a big plus compared to PID type controllers. Owing to the low-footprint form presented here, linear ADRC can now even be implemented with a computational burden comparable to the discrete-time PID family, depriving them of one of their last distinctive advantages.

% -----------------------------------------------------------------------------
% Section: Conclusions
% -----------------------------------------------------------------------------

\section{Conclusions}

This article introduced an efficient implementation of discrete-time linear active disturbance rejection control (ADRC) with focus on a low computational footprint, thus paving the way for using ADRC in applications with high sampling frequencies, strict timing requirements, or low-cost target processors. Such constraints can, for example, be found in practice in the domain of digital control of power electronics systems \cite{Maksimovic:2004}, with sampling frequencies well above 100\,kHz \cite{Corradini:2015}. For a software-based controller implementation, every arithmetic operation becoming redundant will have a measureable impact on the attainable performance in this application area.

The proposed approach is based on a feedback transfer function representation of ADRC. It combines the best of both worlds of the existing state-space and transfer function implementations: a minimum number of storage variables and less arithmetic operations than any existing implementation. At the same time, all properties of the observer in ADRC are preserved. Most importantly for practical applications, the proposed implementation can be combined with arbitrary control signal constraints (e.\,g.\ rate limitation) without having to worry about additional anti-windup measures.

The improvements on the implementation side of ADRC introduced here should make ADRC an even more compelling choice for solving real-world control problems.

% -----------------------------------------------------------------------------
% -----------------------------------------------------------------------------

\AddToShipoutPicture*{%
  \AtPageLowerLeft{%
    \setlength\unitlength{1cm}%
    \put(1.9,1.8){\begin{minipage}[b]{17.79cm}
    \footnotesize\textcolor{black!50}{%
    \rule{17.79cm}{0.5pt}\vspace{1ex}
    \textbf{\textsc{Erratum}}\\[1ex]
    In the original publication, available at \href{https://doi.org/10.23919/ecc54610.2021.9655120}{https://doi.org/10.23919/ecc54610.2021.9655120}, the two equations for $C_u(z)$ and $C_y(z)$ above equation \refEq{eqn:CuCy} contain a minor mistake. The correct equations are
    \\[1ex]
    $\displaystyle
        C_u(z) =
        \frac{z^{-1}}{b_0}
        \cdot
        \frac{
        \begin{pmatrix} \Vector{k}^\Transpose  &  1 \end{pmatrix}
        \cdot
        \operatorname{adj} \left( \Matrix{I} - z^{-1} \Matrix{A}_\mathrm{ESO} \right)
        \cdot
        \Vector{b}_\mathrm{ESO}
        }{ \left( 1 - z^{-1} z_\mathrm{ESO} \right)^{n+1} }
    $\quad and\quad
    $\displaystyle
        C_y(z) =
        \frac{1}{b_0}
        \cdot
        \frac{
        \begin{pmatrix} \Vector{k}^\Transpose  &  1 \end{pmatrix}
        \cdot
        \operatorname{adj} \left( \Matrix{I} - z^{-1} \Matrix{A}_\mathrm{ESO} \right)
        \cdot
        \Vector{l}
        }{ \left( 1 - z^{-1} z_\mathrm{ESO} \right)^{n+1} }
    $\\[2ex]
    instead of the previously reported
    \\[1ex]
    $\displaystyle
        C_u(z) =
        \frac{z^{-1}}{b_0}
        \cdot
        \frac{
        \begin{pmatrix} \Vector{k}^\Transpose  &  1 \end{pmatrix}
        \cdot
        \operatorname{adj} \left( \Matrix{I} - z^{-1} \Matrix{A}_\mathrm{ESO} \right)^{-1}
        \cdot
        \Vector{b}_\mathrm{ESO}
        }{ \left( 1 - z^{-1} z_\mathrm{ESO} \right)^{n+1} }
    $\quad and\quad
    $\displaystyle
        C_y(z) =
        \frac{1}{b_0}
        \cdot
        \frac{
        \begin{pmatrix} \Vector{k}^\Transpose  &  1 \end{pmatrix}
        \cdot
        \operatorname{adj} \left( \Matrix{I} - z^{-1} \Matrix{A}_\mathrm{ESO} \right)^{-1}
        \cdot
        \Vector{l}
        }{ \left( 1 - z^{-1} z_\mathrm{ESO} \right)^{n+1} }
    $.\\[2ex]
    The author would like to thank Rafa\l{} Mado\'{n}ski, Hu Ke, and Zheng Yaojun for discovering and reporting this issue.
    }\end{minipage}}%
  }
}

% -----------------------------------------------------------------------------
% -----------------------------------------------------------------------------


\begin{thebibliography}{10}
\bibitem{Han:2009}
J.~Han, ``From {PID} to active disturbance rejection control,'' \emph{IEEE
  Transactions on Industrial Electronics}, vol.~56, no.~3, pp. 900--906, 2009.

\bibitem{Gao:2003}
Z.~Gao, ``Scaling and bandwidth-parameterization based controller tuning,'' in
  \emph{Proceedings of the 2003 American Control Conference}, 2003, pp.
  4989--4996.

\bibitem{Zheng:2010}
Q.~Zheng and Z.~Gao, ``On practical applications of active disturbance
  rejection control,'' in \emph{Proceedings of the 29th Chinese Control
  Conference}, 2010, pp. 6095--6100.

\bibitem{Zheng:2018}
------, ``Active disturbance rejection control: Some recent experimental and
  industrial case studies,'' \emph{Control Theory and Technology}, vol.~16,
  no.~4, pp. 301--313, 2018.

\bibitem{Herbst:2020}
G.~Herbst, A.-J. Hempel, T.~G\"{o}hrt, and S.~Streif, ``Half-gain tuning for
  active disturbance rejection control,'' in \emph{Proceedings of the 21st IFAC
  World Congress}, 2020, pp. 1341--1346.

\bibitem{Miklosovic:2006}
R.~Miklosovic, A.~Radke, and Z.~Gao, ``Discrete implementation and
  generalization of the extended state observer,'' in \emph{Proceedings of the
  2006 American Control Conference}, 2006, pp. 2209--2214.

\bibitem{Madonski:2015}
R.~Mado\'{n}ski, Z.~Gao, and K.~\L{}akomy, ``Towards a turnkey solution of
  industrial control under the active disturbance rejection paradigm,'' in
  \emph{2015 54th Annual Conference of the Society of Instrument and Control
  Engineers of Japan (SICE)}, 2015, pp. 616--621.

\bibitem{Herbst:2016a}
G.~Herbst, ``Practical active disturbance rejection control: Bumpless transfer,
  rate limitation, and incremental algorithm,'' \emph{IEEE Transactions on
  Industrial Electronics}, vol.~63, no.~3, pp. 1754--1762, 2016.

\bibitem{Madonski:2015b}
R.~Mado\'{n}ski and P.~Herman, ``Survey on methods of increasing the efficiency
  of extended state disturbance observers,'' \emph{ISA Transactions}, vol.~56,
  pp. 18--27, 2015.

\bibitem{Herbst:Preprint2020}
G.~Herbst, ``Transfer function analysis and implementation of active
  disturbance rejection control,'' 2020,
  \href{http://arxiv.org/abs/2011.01044}{arXiv:2011.01044}.

\bibitem{Xu:2014}
R.~Xu, C.~Zhao, and J.~Xiong, ``An efficient discrete form of active
  disturbance rejection controller,'' in \emph{Proceedings of the 33rd Chinese
  Control Conference}, 2014, pp. 3852--3856.

\bibitem{Herbst:2013}
G.~Herbst, ``A simulative study on active disturbance rejection control
  ({ADRC}) as a control tool for practitioners,'' \emph{Electronics}, vol.~2,
  no.~3, pp. 246--279, 2013.

\bibitem{Oppenheim:2010}
A.~V. Oppenheim and R.~W. Schafer, \emph{Discrete-Time Signal Processing},
  3rd~ed.\hskip 1em plus 0.5em minus 0.4em\relax Prentice Hall Press, 2010.

\bibitem{Suntio:2018}
T.~Suntio, ``On dynamic modeling of {PCM}-controlled converters---buck
  converter as an example,'' \emph{IEEE Transactions on Power Electronics},
  vol.~33, no.~6, pp. 5502--5518, 2018.

\bibitem{Hallworth:2012}
M.~Hallworth and S.~A. Shirsavar, ``Microcontroller-based peak current mode
  control using digital slope compensation,'' \emph{IEEE Transactions on Power
  Electronics}, vol.~27, no.~7, pp. 3340--3351, 2012.

\bibitem{Maksimovic:2004}
D.~Maksimovi\'{c}, R.~Zane, and R.~Erickson, ``Impact of digital control in
  power electronics,'' in \emph{2004 Proceedings of the 16th International
  Symposium on Power Semiconductor Devices and ICs}, 2004, pp. 13--22.

\bibitem{Corradini:2015}
L.~Corradini, D.~Maksimovi\'{c}, P.~Mattavelli, and R.~Zane, \emph{Digital
  Control of High-Frequency Switched-Mode Power Converters}, ser. IEEE Press
  Series on Power Engineering.\hskip 1em plus 0.5em minus 0.4em\relax Wiley,
  2015.
\end{thebibliography}
\end{document}